\documentclass[twocolumn,showpacs,preprintnumbers,amsmath,amssymb]{revtex4}

\usepackage{graphicx}
\usepackage{dcolumn}
\usepackage{bm}
\usepackage{color}

\newcommand{\ket}[1]{|#1\rangle}

\begin{document}

\title{On two-qubit states ordering with quantum discords}
\author{Ma{\l}gorzata Okrasa}
 \email{okrasa@merlin.phys.uni.lodz.pl}
\author{Zbigniew Walczak}
 \email{walczak@merlin.phys.uni.lodz.pl}
\affiliation{%
Department of Theoretical Physics, University of Lodz\\
Pomorska 149/153, 90-236 {\L}\'od\'z, Poland}

\date{\today}

\begin{abstract}
The counterintuitive effect of non-unique 
ordering of two-qubit states with quantum entanglement measures was 
discovered over ten years ago. 
More precisely, it was shown by Monte Carlo simulations that 
there exist states for which the entanglement of formation 
and the negativity do not impose~the~same~ordering of states,
i.e. $E_{F}(\rho_{AB}) \leq (\geq) E_{F}(\rho_{AB}^{\prime})$ 
is not equivalent to
$N(\rho_{AB}) \leq (\geq) N(\rho_{AB}^{\prime})$.
Recently, it was discovered that quantum discord 
and the geometric quantum discord do not necessarily imply 
the same ordering of two-qubit $X$-states, 
which means that the lack of the unique ordering of states with 
quantum entanglement measures goes beyond entanglement.   
Inspired by this observation, we study the problem of the states
ordering with quantum discords, 
considering two-qubit Bell-diagonal states for analytical simplicity.
In particular, we identify some classes of states for which the
states ordering with quantum discords is preserved as long as the
states belong to the same class and give a few illustrative examples.
\end{abstract}


\maketitle

\section{Introduction}
In quantum information theory, the problem of characterization 
of correlations present in a quantum state has been intensively 
studied during the last two decades 
(for review, see \cite{Horodeccy09, Guhne09}).
The most significant progress  has been made in this subject 
in the framework of paradigm based on the entanglement-separability 
dichotomy introduced by Werner \cite{Werner89}.
Within this paradigm, the quantum correlations are identified 
with entanglement which is quantified by many entanglement measures. 
However, it has become clear gradually that  quantum correlations
cannot be only limited to entanglement because separable quantum
states can also have non-classical correlations \cite{Knill98,
Braunstein99, Bennett99, Meyer00, Biham04, Datta05, Datta07}, 
and therefore it has become clear that the entanglement-separability 
paradigm is too narrow and needs reconsideration. 

The first step in this direction was the introduction of the concept
of quantum discord, the difference of two natural extensions of 
the classical mutual information, which can be used as a measure of 
non-classical correlations beyond quantum entanglement 
\cite{Zurek00, Ollivier01}. 

After the recent discovery \cite{Datta08, Lanyon08, Datta09} 
that non-classical correlations other than entanglement can be 
responsible for the quantum computational efficiency of deterministic 
quantum computation with one pure qubit \cite{Knill98},
quantum discord became a subject of intensive study in different
contexts, such as 
complete positivity of reduced quantum dynamics 
\cite{Rodriguez08, Shabani09}, 
broadcasting of quantum states \cite{Piani08},
random quantum states \cite{Ferraro10},
dynamics of quantum discord  
\cite{Werlang09, Maziero09, Fanchini10,  Maziero10, Wang10, Hu11}, 
operational interpretation of quantum discord 
\cite{Madhok11, Cavalcanti11}, 
connection between quantum discord and entanglement
irreversibility \cite{Fanchini11B},
relation between quantum discord and 
distillable entanglement \cite{Streltsov11},
relation between quantum discord and 
distributed entanglement of formation \cite{Fanchini11, Fanchini11C},
and monogamy of quantum discord \cite{Prabhu11, Giorgi11}.
 
For pure states, quantum correlations characterized by 
quantum discord can be identified with quantum entanglement 
as measured by the entanglement of formation. 
However, for mixed states it was shown that quantum entanglement, 
as measured by the entanglement of formation, 
may be smaller or larger than quantum discord 
\cite{Luo08, Ali10, Qasimi11}.

Because evaluation of quantum discord involves 
the complicated optimization procedure,
another measure of non-classical correlations beyond quantum 
entanglement was needed.     
Recently, Daki\'c {\em et al.} \cite{Dakic10} introduced the geometric
quantum discord, which involves a simpler optimization procedure than 
quantum discord.   
The geometric quantum discord was studied in different
contexts, such as the quantum computational efficiency of deterministic 
quantum computation with one pure qubit \cite{Dakic10},
dynamics of the geometric quantum discord  
\cite{Lu10, Altintas10, Yeo10, Bellomo11, Li11}, 
relation between the geometric quantum discord and 
other measures of non-classical correlations 
\cite{Batle11, Qasimi11, Girolami11, Girolami11B}.

The geometric quantum discord could be used 
instead of quantum discord, if one showed that they 
give consistent results, taking into account that 
both quantum discords are observable measures of quantum 
correlations \cite{Zhang11, Girolami11C}.

Recently, Yeo {\em et al.} \cite{Yeo10} discovered, quite
unexpectedly, that quantum discords do not give consistent results,
namely they showed that quantum discord and the geometric quantum 
discord do not necessarily imply the same ordering of two-qubit 
$X$-states which means that the following condition 
\begin{align}
\label{OR}
{\cal D}_{A}(\rho_{AB}) \leq(\geq) 
{\cal D}_{A}(\rho_{AB}^{\prime}) 
\iff
{\cal D}_{A}^{G}(\rho_{AB}) \leq(\geq) 
{\cal D}_{A}^{G}(\rho_{AB}^{\prime})
\end{align}
is not satisfied for arbitrary states $\rho_{AB}$
and $\rho_{AB}^{\prime}$.
Therefore, the lack of the unique ordering of states with 
quantum entanglement measures 
\cite{Eisert99, Zyczkowski99, Virmani00, Zyczkowski02, 
Wei03, Miranowicz04, Miranowicz04B, Miranowicz04C, Miranowicz04D, 
Ziman06, Kinoshita07}
goes beyond entanglement.

In this paper, we investigate the problem of the states ordering 
with quantum discords considering a large family of two-qubit states, 
namely two-qubit Bell-diagonal states.
In particular, we identify wide classes of states for which
quantum discords give consistent results.

\section{Quantum discord and geometric quantum discord} 
The quantum mutual information of a state $\rho_{AB}$,
\begin{equation} 
\label{QMI}
{\cal I}(\rho_{AB}) = S(\rho_{A}) + S(\rho_{B}) - S(\rho_{AB}),
\end{equation}
is a measure of the total correlations present in a state $\rho_{AB}$, 
where $\rho_{A(B)}$ is the reduced state of the system $A(B)$, 
and $S(\rho) = - \text{Tr}(\rho \log_{2} \rho)$ is the von Neumann
entropy. 
The quantum conditional entropy, 
$S(\rho_{B|A}) = S(\rho_{AB}) - S(\rho_{A})$, 
allows one to rewrite the quantum mutual information 
in the following form 
\begin{equation}
\label{QMI2}
{\cal I}(\rho_{AB}) = S(\rho_{B}) - S(\rho_{B|A}).
\end{equation}  
The fact that the quantum conditional entropy quantifies 
the ignorance about the system $B$ that remains if we make
measurements on the system $A$ allows one to find an alternative 
expression for the quantum conditional entropy, and thereby for 
the quantum mutual information.

If the von Neumann projective measurement, 
described by a complete set of one-dimensional
orthogonal projectors, $\{\Pi_{i}^{A}\}$, corresponding 
to outcomes $i$, is performed, then the state of the system $B$ 
after the measurement is given by 
$\rho_{B|i} = 
\text{Tr}_{A}[(\Pi_{i}^{A} \otimes I) \rho_{AB}
              (\Pi_{i}^{A} \otimes I)]/p_{i}^{A}$, 
where $p_{i}^{A} = \text{Tr}[(\Pi_{i}^{A} \otimes I) \rho_{AB}]$.
The von Neumann entropies $S(\rho_{B|i})$, weighted by probabilities 
$p_{i}^{A}$, lead to the quantum conditional entropy of the system
$B$ given the complete measurement $\{\Pi_{i}^{A}\}$ on the system $A$
\begin{equation}
S_{\{\Pi_{i}^{A}\}}(\rho_{B|A}) = \sum_{i} p_{i}^{A} S(\rho_{B|i}),
\end{equation} 
and thereby the quantum mutual information, induced by 
the von Neumann measurement performed on the system $A$, is defined by
\begin{equation}
{\cal J}_{\{\Pi_{i}^{A}\}}(\rho_{AB}) = 
S(\rho_{B}) - S_{\{\Pi_{i}^{A}\}}(\rho_{B|A}).
\end{equation} 
The measurement independent quantum mutual information 
${\cal J}_{A}(\rho_{AB})$, interpreted as a measure of classical
correlations \cite{Henderson01, Ollivier01}, is defined by
\begin{align}
{\cal J}_{A}(\rho_{AB}) & = \sup_{\{\Pi_{i}^{A}\}} 
{\cal J}_{\{\Pi_{i}^{A}\}}(\rho_{AB}). 
\end{align}

In general case, ${\cal I}(\rho_{AB})$ and ${\cal J}_{A}(\rho_{AB})$ 
may differ and the difference, interpreted as a measure 
of quantum correlations, is called quantum discord  \cite{Ollivier01} 
\label{DA}
\begin{align}
{\cal D}_{A}(\rho_{AB}) = 
{\cal I}(\rho_{AB}) - {\cal J}_{A}(\rho_{AB}).
\end{align}
Despite the simplicity of this definition, 
the analytical expressions for quantum discord are known only for
two-qubit Bell-diagonal states \cite{Luo08}, 
for seven-parameter two-qubit $X$-states \cite{Ali10}, 
for two-mode Gaussian states \cite{Giorda10, Adesso10},
and for a class of two-qubit states with parallel nonzero Bloch
vectors \cite{Li11B}. 

Since evaluation of quantum discord involves 
the complicated optimization procedure,
another measure of non-classical correlations beyond quantum 
entanglement was needed.
Recently, Daki\'c {\em et al.} introduced 
the geometric quantum discord \cite{Dakic10} 
\label{GDA}
\begin{align}
{\cal D}_{A}^{G}(\rho_{AB}) = 
\inf_{\chi_{AB}} || \rho_{AB} - \chi_{AB} ||^2,
\end{align}
where the infimum is over all zero-discord states, 
${\cal D}_{A}(\chi_{AB}) = 0$, and 
$|| \cdot ||$ is the Hilbert--Schmidt norm.  
Since the geometric quantum discord involves a simpler 
optimization procedure than quantum discord, 
the analytical expression for geometric quantum discord was obtained 
for arbitrary  two-qubit states \cite{Dakic10} 
as well as for arbitrary bipartite states \cite{Hassan12}.

Recently, Luo and Fu \cite{Luo10} showed that 
the geometric quantum discord is a measurement-based 
measure of non-classical correlations 
closely related to quantum discord.

\section{Two-qubit Bell-diagonal states ordering}
Recently, Yeo {\em et al.} \cite{Yeo10} discovered,
studying the non-Markovian effects  on quantum-communication  
protocols, an explicit example of two two-qubit $X$-states for which 
the states ordering with respect to quantum discord significantly 
differs from that given by the geometric quantum discord.  
More recently, Girolami and Adesso \cite{Girolami11} and 
independently Batle {\em et al.} \cite{Batle11}
provided a numerical comparison between quantum discord and 
the geometric quantum discord for general two-qubit states, 
from which one can infer that there exist other states violating 
the states ordering with quantum discords. 

Inspired by this observation, we investigate the problem of  
the states ordering with quantum discords. 

\begin{figure}
   \centering
   \includegraphics[width=0.49\textwidth]{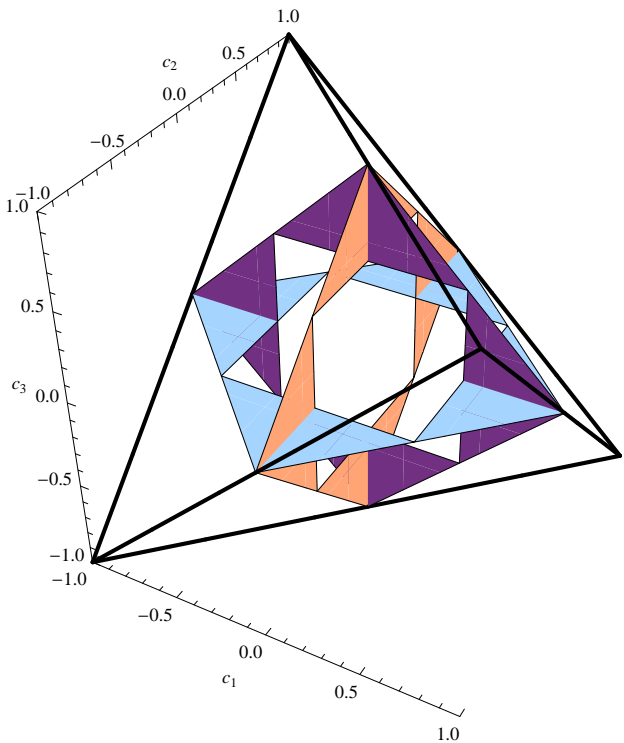}
    \caption{Two-qubit Bell-diagonal states belong to the tetrahedron,
    each of the twelve triangles is a set of states for which the
    states ordering is preserved.} 
   \label{fig1}
\end{figure}

For analytical simplicity, let us consider two-qubit Bell-diagonal
states \cite{Luo08}
\begin{align}
\label{BDS}
\rho_{AB} = \frac{1}{4}(I \otimes I + 
\sum_{i=1}^{3} c_{i}\, \sigma_{i} \otimes \sigma_{i}),
\end{align}
where matrices $\sigma_{i}$ are the Pauli spin matrices 
and real numbers $c_{i}$ fulfill the following conditions
\begin{subequations}
\begin{align}
& 0 \leq \frac{1}{4} (1 - c_{1} - c_{2} - c_{3}) \leq 1, \\
& 0 \leq \frac{1}{4} (1 - c_{1} + c_{2} + c_{3}) \leq 1, \\
& 0 \leq \frac{1}{4} (1 + c_{1} - c_{2} + c_{3}) \leq 1, \\
& 0 \leq \frac{1}{4} (1 + c_{1} + c_{2} - c_{3}) \leq 1.
\end{align}
\end{subequations} 
The above inequalities describe a tetrahedron with vertices
$(1,1,-1)$, $(-1,-1,-1)$,  $(1,-1,1)$ and  $(-1,1,1)$ 
(see Fig.~\ref{fig1}) corresponding to the Bell states
\begin{subequations}
\begin{align} 
& \ket{\psi^{+}} = \frac{1}{\sqrt{2}}(\ket{01} + \ket{10}),\\
& \ket{\psi^{-}} = \frac{1}{\sqrt{2}}(\ket{01} - \ket{10}),\\ 
& \ket{\phi^{+}} = \frac{1}{\sqrt{2}}(\ket{00} + \ket{11}),\\
& \ket{\phi^{-}} = \frac{1}{\sqrt{2}}(\ket{00} - \ket{11}),
\end{align}
\end{subequations}
respectively.

For two-qubit Bell-diagonal states, the quantum discord is given by  
\cite{Luo08}
\begin{align}
\label{QD}
{\cal D}_{A}(\rho_{AB}) & = \frac{1}{4} 
[(1 - c_{1} - c_{2} - c_{3}) \log_{2}(1 - c_{1} - c_{2} - c_{3})
\nonumber\\
& + (1 - c_{1} + c_{2} + c_{3}) \log_{2} (1 - c_{1} + c_{2} + c_{3})
\nonumber\\
& + (1 + c_{1} - c_{2} + c_{3}) \log_{2} (1 + c_{1} - c_{2} + c_{3})
\nonumber\\
& + (1 + c_{1} + c_{2} - c_{3}) \log_{2} (1 + c_{1} + c_{2} - c_{3})]
\nonumber\\
& - \frac{1}{2} [ (1 - c) \log_{2} (1 - c) + (1 + c) \log_{2} (1 + c)], 
\end{align}   
with $c = \max(|c_{1}|, |c_{2}|, |c_{3}|)$, whereas the geometric quantum
discord is given by \cite{Dakic10}
\begin{align}
\label{GQD}
{\cal D}_{A}^{G}(\rho_{AB}) & = \frac{1}{4} 
(c_{1}^{2} + c_{2}^{2} + c_{3}^{2} - c^2). 
\end{align}

It can be verified that for two-qubit Bell-diagonal states, 
if  any two such states 
$\rho_{AB}$ and $\rho_{AB}^{\prime}$
belong to one of twelve triangles 
(see Fig.~\ref{fig1}) with vertices 
\begin{subequations}
\label{tri}
\begin{align}
& (0, -1, 0), && (0, -0.5, 0.5), && (0, -0.5,-0.5) \label{tri1}\\
& (0, 1, 0),  && (0, 0.5, 0.5),  && (0, 0.5, -0.5)\\
& (0, 0, -1), && (0, -0.5, -0.5),&& (0, 0.5, -0.5)\\
& (0, 0, 1),  && (0, 0.5, 0.5),  && (0, -0.5, 0.5)\\
& (-1, 0, 0), && (-0.5, 0, 0.5), && (-0.5, 0, -0.5)\\
& (1, 0, 0),  && (0.5, 0, 0.5),  && (0.5, 0, -0.5)\\
& (0, 0, -1), && (0.5, 0, -0.5), && (-0.5, 0, -0.5) \label{tri7}\\
& (0, 0, 1),  && (0.5, 0, 0.5),  && (-0.5, 0, 0.5) \label{tri8}\\
& (-1, 0, 0), && (-0.5, 0.5, 0), && (-0.5, -0.5, 0)\\
& (1, 0, 0),  && (0.5, -0.5, 0), && (0.5, 0.5, 0)\\
& (0, -1, 0), && (0.5, -0.5, 0), && (-0.5, -0.5, 0)\\
& (0, 1, 0),  && (0.5, 0.5, 0),  && (-0.5, 0.5, 0)
\end{align}
\end{subequations}
then the states ordering (\ref{OR}) is preserved, 
otherwise one can find states for which 
it is violated. 

In other words, if $\rho_{AB}$ and $\rho_{AB}^{\prime}$ 
belong to one of twelve two-parameter families of states 
corresponding to triangles (\ref{tri})
\begin{subequations}
\label{fam}
\begin{align}
c_{1} = 0, \quad\quad & -1 \leq c_{2} \leq -0.5, 
           \quad & |c_{3}| \leq 1 + c_{2}\label{fam1}\\
c_{1} = 0, \quad\quad & 0.5 \leq c_{2} \leq 1, 
           \quad & |c_{3}| \leq 1 - c_{2}\\
c_{1} = 0, \quad\quad & -1 \leq c_{3} \leq -0.5, 
           \quad & |c_{2}| \leq 1 + c_{3} \\
c_{1} = 0, \quad\quad & 0.5 \leq c_{3} \leq 1, 
           \quad & |c_{2}| \leq 1 - c_{3} \\
c_{2} = 0, \quad\quad &  -1 \leq c_{1} \leq -0.5, 
           \quad & |c_{3}| \leq 1 + c_{1} \\
c_{2} = 0, \quad\quad & 0.5 \leq c_{1} \leq 1, 
           \quad & |c_{3}| \leq 1 - c_{1} \\
c_{2} = 0, \quad\quad & -1 \leq c_{3} \leq -0.5, 
           \quad & |c_{1}| \leq 1 + c_{3} \label{fam7}\\
c_{2} = 0, \quad\quad & 0.5 \leq c_{3} \leq 1, 
           \quad & |c_{1}| \leq 1 - c_{3} \label{fam8}\\
c_{3} = 0, \quad\quad & -1 \leq c_{1} \leq -0.5, 
           \quad & |c_{2}| \leq 1 + c_{1} \\
c_{3} = 0, \quad\quad & 0.5 \leq c_{1} \leq 1, 
           \quad & |c_{2}| \leq 1 - c_{1} \\
c_{3} = 0, \quad\quad & -1 \leq c_{2} \leq -0.5, 
           \quad & |c_{1}| \leq 1 + c_{2} \\
c_{3} = 0, \quad\quad & 0.5 \leq c_{2} \leq 1, 
           \quad & |c_{1}| \leq 1 - c_{2}
\end{align}
\end{subequations}
then the states ordering (\ref{OR}) is preserved
otherwise, one can find both the states for which  the states 
ordering is violated and the states for which it is preserved.
A few illustrative examples are given below.

\subsection{Example 1}
\begin{figure}
   \centering
   \includegraphics[width=0.49\textwidth]{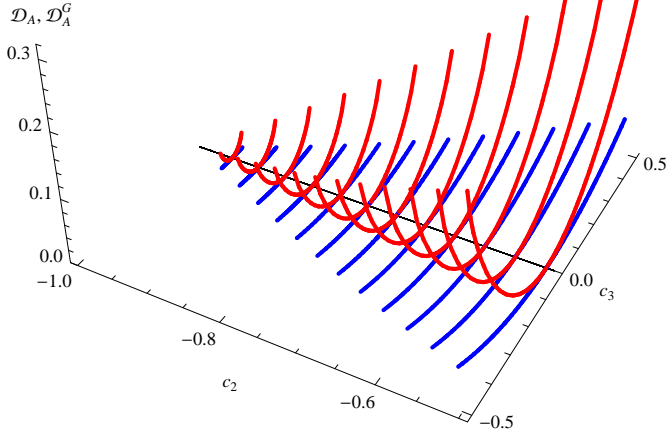}
    \caption{Comparison of quantum discord (red line) and 
             the geometric quantum discord (blue line) 
             for two-qubit Bell-diagonal states with
             $c_{1} = 0, \quad -1 \leq c_{2} \leq -0.5, \quad 
             |c_{3}| \leq 1 + c_{2}$, for selected values of $c_{2}$.} 
   \label{fig2}
\end{figure}

Let us consider states $\rho_{AB}$ and $\rho_{AB}^{\prime}$
belonging to the triangle with vertices (\ref{tri1}), 
i.e. states belonging to the two-parameter family 
(\ref{fam1}).
It can be shown via equations (\ref{QD}) and (\ref{GQD}) that for
these states we have  
\begin{subequations}
\label{discords}
\begin{align}
{\cal D}_{A}(\rho_{AB}) & = \frac{1}{4} 
[(1 - c_{2} - c_{3}) \log_{2}(1 - c_{2} - c_{3})
\nonumber\\
& + (1 + c_{2} + c_{3}) \log_{2} (1 + c_{2} + c_{3})
\nonumber\\
& + (1 - c_{2} + c_{3}) \log_{2} (1 - c_{2} + c_{3})
\nonumber\\
& + (1 + c_{2} - c_{3}) \log_{2} (1 + c_{2} - c_{3})]
\nonumber\\
&  - \frac{1}{2} [ (1 - c_{2}) \log_{2} (1 - c_{2}) 
\nonumber\\
& + (1 + c_{2}) \log_{2} (1 + c_{2})],\\
{\cal D}_{A}^{G}(\rho_{AB}) & = \frac{1}{4} c_{3}^{2}.
\end{align}
\end{subequations}
Let us note that for the states (\ref{fam1})
the ordering of states is preserved 
because for given $c_{2}$ quantum discords (\ref{discords})
are convex functions of $c_{3}$ with local minimum at the same point
and ${\cal D}_{A}(\rho_{AB}) \geq {\cal D}_{A}^{G}(\rho_{AB})$
(see Fig.~\ref{fig2}). In the similar way, 
it can be shown that the ordering of states is preserved
in the case of other families of states (\ref{fam}).

\subsection{Example 2}
Let us consider two states, namely $\rho_{AB}$ with
\begin{align}
c_{1} = 0.1, \quad c_{2} = 0, \quad c_{3} = -0.75
\end{align}
and $\rho_{AB}^{\prime}$ with
\begin{align}
c_{1} = 0.1, \quad  c_{2} = 0, \quad c_{3} = 0.9.
\end{align} 
It can be verified via equations (\ref{fam7}) and
(\ref{fam8}) that $\rho_{AB}$ belongs to the triangle with vertices
(\ref{tri7}), while $\rho_{AB}^{\prime}$ belongs to the triangle with
vertices (\ref{tri8}).
It can be shown directly via equations (\ref{QD}) and (\ref{GQD}) that for
these states the ordering of states is violated because
\begin{subequations}
\begin{align}
{\cal D}_{A}(\rho_{AB}) & \simeq 0.0169,\\
{\cal D}_{A}^{G}(\rho_{AB}) & = 0.0025,
\end{align}
\end{subequations}
and
\begin{subequations}
\begin{align}
{\cal D}_{A}(\rho_{AB}^{\prime}) & \simeq 0.0519,\\
{\cal D}_{A}^{G}(\rho_{AB}^{\prime}) & = 0.0025.
\end{align}
\end{subequations}

\subsection{Example 3}
\begin{figure}
   \centering
   \includegraphics[width=0.49\textwidth]{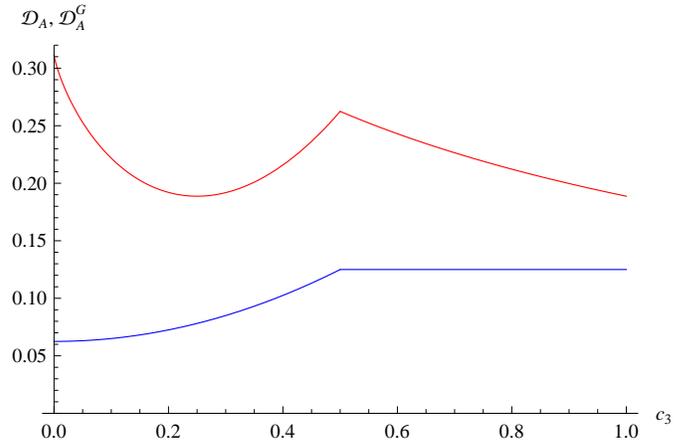}
    \caption{Comparison of quantum discord (red line) 
            and the geometric quantum discord (blue line) 
            for two-qubit Bell-diagonal states with 
            $c_{1} = -0.5, \quad c_{2}= 0.5, \quad 0 < c_{3} \leq 1$.} 
   \label{fig3}
\end{figure}

Let us consider a one-parameter family of states with
\begin{align}
c_{1} = -0.5, \quad c_{2}= 0.5, \quad 0 < c_{3} \leq 1.
\end{align}
It can be verified that these states do not belong to any of triangles
with vertices (\ref{tri}), i.e. families of states
(\ref{fam}), 
and moreover it can be shown via equations 
(\ref{QD}) and (\ref{GQD}) that for these states the ordering of
states is violated (see Fig.~\ref{fig3}).

\subsection{Example 4}
\begin{figure}
   \centering
   \includegraphics[width=0.49\textwidth]{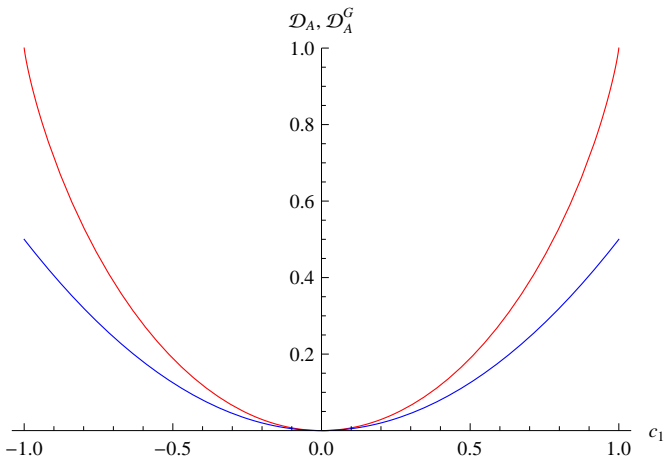}
    \caption{Comparison of quantum discord (red line) 
            and the geometric quantum discord (blue line) 
            for two-qubit Bell-diagonal states with 
            $c_{1} \neq 0, \quad c_{2}= -c_{1}, \quad c_{3} = 1$.}
   \label{fig4}
\end{figure}

Let us consider a one-parameter family of states with
\begin{align}
c_{1} \neq 0, \quad c_{2} = -c_{1}, \quad c_{3} = 1.
\end{align}
It can be verified that these states do not belong to any of triangles
with vertices (\ref{tri}), i.e. families of states (\ref{fam}), 
and moreover it can be shown via equations 
(\ref{QD}) and (\ref{GQD}) that for these states the ordering of
states is preserved (see Fig.~\ref{fig4}).

\section{Summary}
We have investigated the problem of the states ordering with 
quantum discord and the geometric quantum discord.
For analytical simplicity, we have considered 
two-qubit Bell-diagonal states. 
We have identified twelve two-parameter families of states 
for which the states ordering with quantum discords is preserved 
as long as the states belong to the same family, and we have found 
that otherwise, one can find both the states 
for which  the states ordering is violated and the states for
which it is preserved. 
We have also shown that counterintuitively the ordering of states 
can be violated in the case of states belonging to different families,
which explains why a general solution of the problem of two-qubit
states ordering with quantum discords is a challenging issue worth 
further investigations. 
Moreover, a few explicit examples have been given to illustrate 
the results.
   
\acknowledgments
This work was supported by the University of Lodz Grant, 
the Polish Ministry of Science and Higher Education 
Grant No. N N202 103738, and the Polish Research Network 
{\em Laboratory of Physical Foundations of Information Processing}. 
MO acknowledges the support from the European Union under 
the Human Capital Programme -- Measure 8.2.1 and the IROP -- Measure 2.6.


\end{document}